

\input phyzzx.tex
\input newphy.tex
\input tables.tex
\def\ZPC#1#2#3{Z. Phys.\ {\bf C#1}, #2 (#3)}
\def\PRL#1#2#3{Phys.\ Rev.\ Lett.\ {\bf #1}, #2 (#3)}
\def\PRD#1#2#3{Phys.\ Rev.\ {\bf D#1}, #2 (#3)}
\def\PLB#1#2#3{Phys.\ Lett.\ {\bf B#1}, #2 (#3)}
\def\PREP#1#2#3{Phys.\ Rep.\ {\bf #1}, #2 (#3)}
\def\NPB#1#2#3{Nucl.\ Phys.\ {\bf B#1}, #2 (#3)}
\def\Re{\mathop{\rm Re}\nolimits}
\def\arcosh{\mathop{\rm arcosh}\nolimits}
\def\hc{{\rm h.c.}}
\def\tev{\,{\rm TeV }}
\def\gev{\,{\rm GeV }}
\def\drbar{{\overline{\rm DR}}}
\def\msbar{{\overline{\rm MS}}}
\def\mgut{M_{\rm GUT}}
\def\s{s_{\rm w}}
\def\c{c_{\rm w}}
\def\as{\alpha_{\rm s}}
\def\d{{\rm d}}
\def\O{{\cal O}}
\date={}
\titlepage
\hoffset=-.65cm
\voffset=-.4cm
\hsize=17.5cm
\vsize=23.cm
\line{\bf KEK--TH--408\hfill}
\line{\bf KEK Preprint 94--74\hfill}
\line{\bf hep-ph/9408313\hfill}
\line{August 1994\hfill}
\vskip1.cm
\title{\bf On the relation between the fermion pole mass and $\msbar$ Yukawa
coupling in the standard model}
\author{Ralf Hempfling$^1$\footnote{*}{Address after 1 October 1994:
{\it Max-Planck-Institut f\"ur Physik, F\"ohringer Ring 6, 80805 Munich,
Germany}.}
and Bernd A. Kniehl$^{2^{\scriptstyle\,*}}$\footnote{\dag}{On leave from
{\it II. Institut f\"ur Theoretische Physik, Universit\"at Hamburg,
Luruper Chaussee 149, 22761 Hamburg, Germany}.}\footnote{\ddag}{JSPS Fellow.}}
\address{$^1$DESY Theory Group, Notkestra\ss e 85, 22603 Hamburg, Germany}
\address{$^2$KEK Theory Group, 1-1 Oho, Ibaraki-ken, Tsukuba 305, Japan}
\vfill
\abstract
We present the full one-loop correction to the relation between the running
$\msbar$ Yukawa coupling, $\overline h_f(\mu)$, at some renormalization scale,
$\mu$, and the pole mass, $m_f$, of a fermion, $f$, in the standard model.
Our result complements previous analyses that included just the QCD
correction to this relation in the case of quarks.
This allows us to convert, without loss of information, the threshold value,
$\overline h_f(m_f)$, of $\overline h_f(\mu)$, which satisfies a
renormalization-group equation and may be driven up to some high-energy scale
of new physics, to physical observables at low energy.
We investigate the limit of the Higgs boson and/or the top quark being much
heavier than all the other particles.
We also list simple and yet rather precise approximation formulae, which are
convenient for applications.

\vskip1.cm
\noindent
PACS number(s): 11.10.Hi, 12.15.Ff, 12.15.Lk
\endpage

\REF\amal{J. Ellis, S. Kelley, and D.V. Nanopoulos, \PLB{260}{131}{1991};
U. Amaldi, W. de Boer, and H. F\"urstenau, \PLB{260}{447}{1991};
P. Langacker and M. Luo, \PRD{44}{817}{1992}.}
\REF\pdg{K. Hikasa \etal\ (Particle Data Group), \PRD{45}{S1}{1992}.}
\REF\suss{See, for example, L. Susskind, \PREP{104}{181}{1984}.}
\REF\dhr{See, for example, S. Dimopoulos, L.J. Hall, and S. Raby,
\PRL{68}{1984}{1992}; \PRD{45}{4192}{1992}.}
\REF\ybeta{H. Arason, D.J. Casta\~no, B. Keszthelyi, S. Mikaelian, E.J. Piard,
P. Ramond, and B.D. Wright, \PRL{67}{2933}{1991}; \PRD{46}{3945}{1992}
and references cited therein.}
\REF\sirlin{A. Sirlin and R. Zucchini, \NPB{266}{389}{1986}.}
\REF\pierce{D. Pierce, Johns Hopkins University Report No.\ JHU-TIPAC-940010,
talk presented at the SUSY-94 Conference, Ann Arbor, Michigan, May 14--17,
1994.}
\REF\reduct{W. Siegel, {Phys.\ Lett.} {\bf 84B}, (1979) 193;
D.M. Capper, D.R.T. Jones, and P. van Nieuwenhuizen,
\NPB{167}{479}{1980}.}
\REF\mssm{H.E. Haber and R. Hempfling, \PRL{66}{1815}{1991}.}
\REF\pole{S. Willenbrock and G. Valencia, \PLB{259}{373}{1991};
R.G. Stuart, \PLB{262}{113}{1991}; {\it ibid.} {\bf B272}, 353 (1991);
\PRL{70}{3193}{1993};
A. Sirlin, \PRL{67}{2127}{1991}; \PLB{267}{240}{1991}.}
\REF\hff{B.A. Kniehl, \NPB{376}{3}{1992}.}
\REF\wjm{W.J. Marciano, private communication.}
\REF\rgs{R.G. Stuart, private communication.}
\REF\ray{A.I. Bochkarev and R.S. Willey, University of Pittsburgh Report No.\
PITT-TH-94-04 (June 1994).}
\REF\sirlinmu{A. Sirlin, \PRD{22}{971}{1980}.}
\REF\hzz{B.A. Kniehl, \NPB{352}{1}{1991}.}
\REF\arbitraryxi{G. Degrassi and A. Sirlin, \NPB{383}{73}{1992}.}
\REF\gra{N. Gray, D.J. Broadhurst, W. Grafe, and K. Schilcher,
\ZPC{48}{673}{1990}.}
\REF\gro{E. Gross, B.A. Kniehl, and G. Wolf,
Report No.\ DESY~94-035 and WIS-94/15/MAR-PH (March 1994),
Z.\ Phys.\ C (to appear).}
\REF\vel{M. Veltman, Acta Phys.\ Pol.\ {\bf B8}, 475 (1977).}
\REF\lep{The LEP Collaborations ALEPH, DELPHI, L3, OPAL and The LEP
Electroweak Working Group, CERN Report No.\ CERN/PPE/93-157 (26 August 1993).}
\REF\blo{A. Blondel, to appear in
{\it Proceedings of the 1994 Zeuthen Workshop on Elementary Particle Theory:
Physics at LEP200 and Beyond}, Teupitz, Germany, April 10--15, 1994,
edited by J. Bl\"umlein and T. Riemann,
Nucl.\ Phys.\ B (Proceedings Supplements).}
\REF\wei{A.J. Weinstein and R. Stroynowski,
Ann.\ Rev.\ Nucl.\ Part.\ Phys.\ {\bf43}, 457 (1993).}
\REF\dom{C.A. Dominguez and N. Paver, \PLB{293}{197}{1992}.}
\REF\cdf{F. Abe \etal\ (CDF Collaboration), Report No.\
FERMILAB-PUB-94/097-E and CDF/ PUB/TOP/PUBLIC/2561 (April 1994),
submitted to Phys.\ Rev.\ D; \PRL{73}{225}{1994}.}
\REF\jeg{F. Jegerlehner, in
{\it Testing the Standard Model---Pro\-ceedings of the 1990 Theoretical
Advanced Study Institute in Elementary Particle Physics},
edited by M. Cveti\v c and P. Langacker (World Scientific, Singapore, 1991)
p.~476.}
\FIG\figone{Definitions of the gauge-boson and fermion self-energies and
the Higgs tadpole contribution.}
\FIG\figtwo{Electroweak threshold corrections to Eq.~{\defymsbar},
$\delta_f^{\rm w}(m_f)+\delta_f^{\rm QED}(m_f)$,
for (a) $f=t$, (b) $f=b$, and (c) $f=\tau$
at one loop in the SM as functions of $M_H$ assuming $m_t=(175\pm25)$~GeV.
In Figs.~(b) and (c), decoupling effects are not corrected for.
The dotted lines correspond to the approximations by Eqs.~{\dtop}--{\dtau}
plus Eq.~{\appr}.}
\FIG\figthree{Electroweak threshold corrections to Eq.~{\defymsbar},
$\delta_f^{\rm eff}(m_f)+\delta_f^{\rm QED}(m_f)$,
for (a) $f=b$, (b) $f=\tau$, and (c) their difference
at one loop in the SM as functions of $M_H$ assuming $m_t=(175\pm25)$~GeV.
In contrast to Figs.~{\figtwo}(b) and (c), decoupling effects are corrected
for.}

\chap{Introduction}

Perhaps, some of the most puzzling questions concerning the standard model (SM)
are related to the origin of the Cabibbo-Kobayashi-Maskawa matrix and
the large ratio of fermion masses, \eg, the ratio of the top-quark mass to the
electron mass is $m_t/m_e\approx 3\times 10^5$.
There have been many attempts to try and explain the
seemingly random low-energy values of the Yukawa matrices
as the result of some underlying structure imposed by a new symmetry
at a higher scale, possibly in the framework of a grand unified theory (GUT).
Of course, the SM GUT is ruled out because
(i) the gauge couplings do not meet within
the experimental errors [\amal],
(ii) the predicted proton lifetime is too low [\pdg],
and (iii) the large hierarchy between the GUT scale, $\mgut$, and the
electroweak scale is unstable under radiative corrections and requires a
very severe fine-tuning [\suss].
However, all these shortcomings can be reconciled
in the minimal supersymmetric extension of the SM (MSSM).

GUT models predict automatically relations amongst the fermion masses.
The simplest and most successful example is the ratio
$m_b/m_\tau$, which can be predicted in SUSY GUTs
assuming Yukawa unification at
$\mgut \approx 2\times 10^{16}$~GeV,
and there have been more ambitious attempts
that tried to incorporate also the first two generations [\dhr].
The dominant radiative corrections to these relations
can be included via renormalization-group (RG)
evolution of the parameters from $\mgut$ to the electroweak scale
using one-loop and two-loop $\beta$ functions [\ybeta].

Other significant effects arise at the thresholds when one
tries to derive a prediction for a physical observable
in terms of the running fundamental parameters.
The QCD portion of these effects is well under control [\ybeta].
The first step to cover also the electroweak part was taken by Sirlin and
Zucchini [\sirlin], who derived the one-loop electroweak correction to the
relation between the pole mass of the SM Higgs boson and its quartic
self-coupling defined in the modified minimal-subtraction
$\left(\msbar\right)$ scheme.
In this paper, we shall extend that analysis to the fermion sector of the SM.
Specifically, we shall compute the full one-loop radiative correction,
$\delta_f(\mu)$, in the relation
$$
\overline h_f(\mu) = 2^{3/4}G_F^{1/2}m_f\left[1+\delta_f(\mu)\right] \,,
\eqn\defymsbar$$
where $G_F=1.166\,39\times10^{-5}$~GeV$^{-2}$ is Fermi's constant [\pdg],
$m_f$ is the pole mass of fermion $f$, and $\overline h_f(\mu)$ is the
running $\msbar$ Yukawa coupling of $f$ at renormalization scale $\mu$.
For the most part our analysis, we shall keep $f$ generic, except for the
numerical discussion, where we shall concentrate on the cases of special
interest, $f=t$, $b$, and $\tau$.
We may view Eq.~{\defymsbar} also as the relation between $\overline h_f(\mu)$
and its counterpart in the on-shell scheme, $h_f=2^{3/4}G_F^{1/2}m_f$.
Notice that, although we focus attention on the effects due to the particles
of the SM,
our analysis can be immediately generalized to any extension thereof.
We note in passing that the one-loop correction to the MSSM relation
$h_f=\sqrt{g^2+g^{\prime2}}\,m_f/\left(\sqrt2M_Z\sin\beta\right)$,
where $g$ and $g^\prime$ are the SU(2)$_L$ and U(1)$_Y$ gauge couplings,
respectively, and $\tan\beta=v_2/v_1$, with $v_1$ and $v_2$ being the
vacuum expectation values (VEVs) of the two Higgs doublets,
has been found recently [\pierce] in the framework of a minimal subtraction
scheme in which the divergences are regularized by dimensional reduction
$\left(\drbar\right)$ [\reduct].
However, this result does not include the solution to our problem,
since, in Ref.~{\pierce}, $g$, $g^\prime$, $v_1$, and $v_2$ are defined as
$\drbar$ parameters, and their expressions in terms of $G_F$ and the physical
particle masses are not specified.

This paper is organized as follows.
In Sect.~2, we shall explain our treatment of the Higgs tadpoles and then
derive the full SM one-loop correction, $\delta_f(\mu)$, in
Eq.~{\defymsbar}.
We shall first assume that $f$ is the heaviest particle of the SM.
In a second step, we shall incorporate the effects of decoupling of weak
physics at intermediate scales, which will eliminate the leading
logarithmic (LL) corrections.
Section~3 contains our numerical discussion.
Our conclusions are summarized in Sect.~4.
Lengthy expressions are relegated to the Appendix.

\chap{Relation between the fermion pole mass and $\msbar$ Yukawa coupling}

To start with, we consider the Higgs sector and describe how we handle
the tadpoles.
The most general gauge-invariant renormalizable potential
of one complex SU(2)$_L$-doublet scalar field, $\Phi$, with $Y = 1$
can be written as
$$
V = {\lambda_0\over2}\left(\Phi^\dagger \Phi\right)^2
- m_0^2 \Phi^\dagger \Phi\,.
\eqno\eq$$
Here and in the following, we denote all bare quantities by a subscript 0.
With the replacement $\Phi \rightarrow (v_0+H)/\sqrt{2}$,
where $v_0$ is the VEV of the CP-even component of $\Phi$,
we obtain
$$
V = t H + {m_H^2\over2} H^2 + \O\left(H^3\right)\,.
\eqno\eq$$
Here, we have defined $t = v_0\left(\lambda_0 v_0^2/2 - m_0^2\right)$
and $m_H^2 = 3 \lambda_0 v_0^2/2 - m_0^2$.
Since $v_0$ is not a physical quantity, its definition
at the one-loop level is arbitrary and a matter of convenience.
There are two obvious choices one could make.
One possibility is to define $v_0$ such that the bare one-point vertex and the
one-loop tadpole contribution, $T$, defined in Fig.~{\figone} cancel, \ie,
$t + T = 0$. This has the advantage that the tadpoles never enter
a calculation that does not involve other observables of the Higgs sector,
such as the mass or the self-coupling of the Higgs boson, at tree level.
Examples for calculations with an explicit dependence on the tadpoles
are presented in Refs.~{\sirlin} and {\mssm}.
Another possibility is to define $v_0 = m_0 \lambda_0^{-1/2}$,
which has the advantage that $v_0$ is expressed only in terms of
bare parameters and is thus manifestly gauge independent.
Since we wish to use gauge independence and the explicit cancellation of the
tadpoles as a welcome check of our calculation,
we shall choose the first definition.

The tree-level relation between the mass and the Yukawa coupling of an
up-type fermion, $U$, is fixed by the Yukawa Lagrangian,
$$
{\cal L} = h_{U0} \overline Q_L \Phi U_R + \hc \supset
m_{U0} \overline U_L U_R + \hc \,,
\eqno\eq$$
and similarly for a down-type fermion, $D$.
Here, the subscripts $L$ and $R$ label left- and right-handed fields,
respectively, and $Q_L = (U_L, D_L)$.
Thus, one has
$$m_{f0} = 2^{-1/2}v_0h_{f0}\,,
\eqn\bare$$
with $f = U, D$.
In the following, we shall incorporate the full one-loop correction to this
relation in the SM.
For applications, it is most useful to express this relation in terms of
$G_F$, $\overline h_f(\mu)$, and $m_f$ [see Eq.~{\defymsbar}].

The pole mass of a particle is defined as the real part of the complex pole
position of its propagator and is a constant of nature [\pole].
In applications, it is often more convenient to deal with the on-shell mass,
which corresponds to the zero of the real part of the inverse propagator.
In the approximation of neglecting the momentum dependence of the imaginary
part of the self-energy, the two mass definitions coincide.
For the purposes of the one-loop analysis carried out here, we need not
distinguish between them.
At one loop, $m_f$ and $m_{f0}$ are related by
$$
m_f = m_{f0} \left\{ 1
- \Re\left[\Sigma_V^f\left(m_f^2\right)+\Sigma_S^f\left(m_f^2\right)\right]
+ 2^{1/4} G_F^{1/2} {T\over M_H^2}\right\}\,,
\eqn\defmtpole$$
where $\Sigma_V^f(q^2)$ and $\Sigma_S^f(q^2)$ are the tadpole-free vector and
scalar components of the self-energy of $f$ at four-momentum $q$,
respectively.
The precise definitions of these amplitudes may be seen from Fig.~{\figone};
analytic expressions in the 't~Hooft-Feynman gauge may be found in Ref.~{\hff}.
For the reader's convenience, we list
$\Sigma_V^f\left(m_f^2\right)+\Sigma_S^f\left(m_f^2\right)$ and $T$ in the
Appendix.
Throughout this work, we use dimensional regularization with $n=4-2\epsilon$
being the dimensionality of space-time.
Adopting the $R_\xi$ gauge, we have verified that the right-hand side of
Eq.~{\defmtpole} is independent of the four SM gauge parameters,
$\xi_W$, $\xi_Z$, $\xi_\gamma$, and $\xi_g$, as it should be.

As far as QCD corrections are concerned, the term $\delta_f(\mu)$ in
Eq.~{\defymsbar} is well known (see, \eg, Ref.~{\ybeta}).
In fact, it is the very correction that appears in the relation between
$m_f$ and the $\msbar$ mass, $\overline m_f(\mu)$, evaluated at
renormalization scale $\mu$, so that
$\overline h_f(\mu)=2^{3/4}G_F^{1/2}\overline m_f(\mu)$.
Here it is understood that only QCD corrections are taken into account.
The purely electromagnetic correction may be inferred from this relation in the
usual way, by substituting $Q_f^2\alpha$ for $C_F\as(\mu)$,
where $Q_f$ is the electric charge of $f$ in units of the positron charge,
$\alpha$ is the fine-structure constant,
$C_F=\left(N_c^2-1\right)/(2N_c)$, with $N_c=3$, is the quadratic Casimir
operator of SU(3)$_c$, and $\as(\mu)$ is the strong coupling constant at scale
$\mu$.
In the case of the weak correction, things are more involved, since
the renormalization of the VEV must be included, too.
As a consequence, in the full SM,  $\overline h_f(\mu)$ will not be
proportional to some $\msbar$ fermion mass to be defined.

Nevertheless, it is interesting in its own right to think about the definition
of $\overline m_f(\mu)$ in the presence of weak interactions---especially,
since we were not able to locate a discussion of this matter in the
literature [\wjm].
By analogy to the QCD case, we may convert Eq.~{\defmtpole} from the on-shell
scheme to the $\msbar$ scheme by keeping just the terms proportional to
$\Delta=1/\epsilon -\gamma_E+\ln(4\pi)$, where $\gamma_E$ is the
Euler-Mascheroni constant, within the curly brackets.
We may then eliminate $m_{f0}$ to obtain a relation between
$\overline m_f(\mu)$ and $m_f$,
$$
\overline m_f(\mu) = m_f \left\{ 1
+ \Re\left[\Sigma_V^f\left(m_f^2\right)+\Sigma_S^f\left(m_f^2\right)\right]
- 2^{1/4} G_F^{1/2} {T\over M_H^2}\right\}_{\Delta=0}\,.
\eqn\defmtms$$
It is interesting to observe that we need to keep the tadpole contribution in
order to obtain a gauge-independent definition of $\overline m_f(\mu)$.
This introduces corrections that grow like $1/M_H^2$ for $M_H$ decreasing,
which seems peculiar, since there is no theoretical principle that prohibits
the Higgs boson from being very light.
On the other hand, if we chose to discard the tadpole contribution in
Eq.~{\defmtms}, $\overline m_f(\mu)$ would depend on $\xi_W$ and $\xi_Z$.
One might argue that this does not create a problem because, in contrast to
$m_f$, $\overline m_f(\mu)$ does not represent a physical observable and is
thus entitled to be gauge dependent [\rgs].
This avenue was taken recently in Ref.~{\ray} for the definition of
$\overline m_t(\mu)$ in the high-$m_t$ approximation.
Since the light-Higgs-boson scenario is now excluded experimentally,
we express a preference for the definition according to Eq.~{\defmtms}.

We now return to our original problem of determining the full one-loop
expression for $\delta_f(\mu)$ in Eq.~{\defymsbar}.
The Fermi constant is not a parameter of the SM Lagrangian.
In order to express it in terms of SM parameters, we need to compute a physical
process both in the Fermi model and the SM and equate the two results.
This was done first in a pioneering paper by Sirlin [\sirlinmu], who considered
the muon lifetime.
In our notation, the result of Ref.~{\sirlinmu} may be written as
$$
G_F = {1\over \sqrt{2}\,v_0^2}
\left[1+{\Pi_{WW}(0)\over M_W^2}-2^{5/4}G_F^{1/2}{T\over M_H^2}+E \right]\,.
\eqn\defgmu$$
Here, $\Pi_{WW}(q^2)$ is the tadpole-free $W$-boson self-energy at
four-momentum $q$, which may be found in Ref.~{\hzz};
for completeness, we list $\Pi_{WW}(0)$ in the Appendix.
$E$ comprises the wave-function renormalizations and the vertex and box
corrections that the SM introduces on top of the calculation of the
muon decay width in the QED-improved Fermi model.
In Ref.~{\sirlinmu}, $E$ is presented in the 't~Hooft-Feynman gauge,
with $\xi_W=\xi_Z=\xi_\gamma=1$,
$$
E={\alpha\over4\pi\s^2}\left[4\left(\Delta-\ln{M_Z^2\over\mu^2}\right)
+\left({7\over2\s^2}-6\right)\ln\c^2+6\right]\,.
\eqno\eq$$
Here and in the following, we use the abbreviation
$\c^2=1-\s^2=M_W^2/M_Z^2$ [\sirlinmu] and identify
$\alpha/\s^2=\sqrt2G_FM_W^2/\pi$ at the one-loop level.
Note that there are no counterterms on the right-hand side of Eq.~{\defgmu},
since the tree-level amplitude is still parametrized in terms of
the bare (and divergent) quantity $v_0$.
An explicit check of the gauge independence of the right-hand side of
Eq.~{\defgmu} would require the calculation of $\Delta r$ [\sirlinmu] in an
arbitrary $R_\xi$ gauge, which is beyond the scope of this work.
However, if we assume that $\Delta r$ is gauge independent then
the gauge independence of Eq.~{\defgmu} follows immediately from the results
of Refs.~{\sirlinmu} and {\arbitraryxi}.

Next, we have to define the renormalized Yukawa coupling.
At one loop in the $\msbar$ scheme, we have
$$
\overline h_f(\mu)=h_{f0} - \delta h_f\,,
\eqn\defhf$$
where the counterterm,
$$
\delta h_f= {\beta_f\over2}\Delta\,,
\eqn\hfct$$
only subtracts the divergent piece proportional to $\Delta$.
Here, $\beta_f$ is the familiar beta function of the SM Yukawa sector
[\ybeta], which may be written in the compact form
$$
\beta_f={h_f\over16\pi^2}\left[{3\over2}\left(h_f^2-h_{f^\prime}^2\right)
+\sum_iN_ih_i^2-{9\over4}g^2-{3\over4}g^{\prime2}
\left(Y_{f_L}^2+Y_{f_R}^2\right)-6C_Fg_{\rm s}^2\right]\,,
\eqn\betaf$$
where $g^2\s^2=g^{\prime2}\c^2=4\pi\alpha$, $g_{\rm s}^2=4\pi\as$,
$f^\prime$ stands for the weak isopartner of $f$, $i$ runs over all
massive fermions, $N_i=1$ (3) for leptons (quarks), and it is understood that
the QCD term is to be omitted if $f$ represents a lepton.
Furthermore, $Y_{f_L}$ and $Y_{f_R}$ denote the weak hypercharges of the
left- and right-handed components of $f$, respectively.
Specifically,
$\left(Y_{e_L},Y_{e_R},Y_{u_L},Y_{u_R},Y_{d_L},Y_{d_R}\right)=
\left(-1,-2,{1\over3},{4\over3},{1\over3},-{2\over3}\right)$ and similarly
for the other generations.
The right-hand side of Eq.~{\defhf} is gauge independent, since $h_{f0}$
is a bare quantity and Eq.~{\hfct} does not carry any gauge parameters.

Inserting Eqs.~{\defmtpole}, {\defgmu}, and {\defhf} into Eq.~{\bare} and
comparing the outcome with Eq.~{\defymsbar}, we obtain the desired result,
$$
\delta_f(\mu)=
\Re\left[\Sigma_V^f\left(m_f^2\right)+\Sigma_S^f\left(m_f^2\right)\right]
-{\Pi_{WW}(0)\over2M_W^2}-{E\over2}-{\beta_f\over2h_f}\Delta\,.
\eqn\delf$$
This is manifestly finite and gauge independent.
Furthermore, we observe that the tadpole contributions have cancelled,
so that there are no inverse powers of $M_H$ left.
Equation~{\delf} may be decomposed into into a weak, an electromagnetic,
and a QCD part, which are separately finite and gauge independent,
$$
\delta_f(\mu)=\delta_f^{\rm w}(\mu)+\delta_f^{\rm QED}(\mu)+
\delta_f^{\rm QCD}(\mu)\,.
\eqno\eq$$
The latter, which is present only if $f$ represents a quark, is quite
familiar [\ybeta],
$$
\delta_f^{\rm QCD}(\mu)=C_F{\as(\mu)\over4\pi}\left(3\ln{m_f^2\over\mu^2}-4
\right)\,,
\eqn\dqcd$$
and even the $\O\left(\as^2\right)$ term is known
[\gra].\footnote{\scriptstyle\#}{A numerically insignificant error in the
relevant formula of Ref.~{\gra} is corrected in Ref.~{\gro}.}
As mentioned before, $\delta_f^{\rm QED}(\mu)$ emerges from Eq.~{\dqcd} by
substituting $Q_f^2\alpha$ for $C_F\as(\mu)$.
$\delta_f^{\rm QCD}(\mu)$ is gauge independent because it is the complete
$\O(\as)$ contribution to the gauge-independent quantity $\delta_f(\mu)$.
This is true also for $\delta_f^{\rm QED}(\mu)$, since it is obtained from
the very set of Feynman diagrams with the gluon replaced by a photon.
Thus, also $\delta_f^{\rm w}(\mu)$ is gauge independent.
In the remainder of this paper, we shall concentrate on
$\delta_f^{\rm w}(\mu)$.

It is interesting to study the limit where $M_H$ and/or $m_t$ are large as
compared to the masses of all the other particles of the SM.
By putting the latter equal to zero, we obtain
$$\eqalignno{
\delta_t^{\rm w}(\mu)&={G_Fm_t^2\over8\pi^2\sqrt2}\left[
-\left(N_c+{3\over2}\right)\ln{m_t^2\over\mu^2}+{N_c\over2}+4-r
+2r(2r-3)\ln(4r)
\vphantom{\left(1-{1\over r}\right)^{3/2}}\right.\cr
&\qquad{}-\left.8r^2\left(1-{1\over r}\right)^{3/2}\arcosh\sqrt r\right]\,,
&\eqnalign\dtop\cr
\delta_b^{\rm w}(\mu)&={G_F\over8\pi^2\sqrt2}\left\{
m_t^2\left[\left(-N_c+{3\over2}\right)\ln{m_t^2\over\mu^2}+{N_c\over2}
-{5\over4}\right]+{M_H^2\over4}\right\}\,,
&\eqnalign\dbot\cr
\delta_f^{\rm w}(\mu)&={G_F\over8\pi^2\sqrt2}\left[
N_cm_t^2\left(-\ln{m_t^2\over\mu^2}+{1\over2}\right)+{M_H^2\over4}\right]\,,
&\eqnalign\dtau\cr}
$$
where $f\ne t,b$.
Equation~{\dtop} is valid for $r=\left(M_H^2/4m_t^2\right)\ge1$.
For $r<1$, one has to replace $(1-1/r)^{3/2}\arcosh\sqrt r$ by
$(1/r-1)^{3/2}\arccos\sqrt r$.
Equation~{\dtop} agrees with Eq.~(15) of Ref.~{\ray}.
Furthermore, we can expand Eq.~{\dtop} for $M_H\gg2m_t$ and $M_H\ll2m_t$,
which yields
$$\eqalignno{
\delta_t^{\rm w}(\mu)&={G_F\over8\pi^2\sqrt2}\left\{{M_H^2\over4}
+m_t^2\left[N_c\left(-\ln{m_t^2\over\mu^2}+{1\over2}\right)
-{3\over2}\ln{M_H^2\over\mu^2}+{7\over4}\right.\right.\cr
&\qquad{}+\left.\left.\O\left({m_t^2\over M_H^2}\ln{M_H^2\over m_t^2}\right)
\right]\right\}\,,
&\eqnalign\heavyh\cr
\delta_t^{\rm w}(\mu)&={G_Fm_t^2\over8\pi^2\sqrt2}\left[
-\left(N_c+{3\over2}\right)\ln{m_t^2\over\mu^2}+{N_c\over2}+4
-2\pi{M_H\over m_t}
+\O\left({M_H^2\over m_t^2}\ln{m_t^2\over M_H^2}\right)\right]\,,
&\eqnalign\heavyt\cr}
$$
respectively.
Equation~{\dtau} emerges from the $W$-boson self-energy and is thus flavour
independent,
while Eqs.~{\dtop} and {\dbot} receive additional contributions from the
top- and bottom-quark self-energies, respectively.
The last term in Eq.~{\delf} just subtracts the ultraviolet divergence,
and $E$ does not survive in the limit of large $M_H$ and/or $m_t$ anyway.
{}From Eqs.~{\dbot}--{\heavyh} we see that a heavy Higgs boson
induces through quantum effects power corrections quadratic in $M_H$.
This is not surprising when we recall that the screening theorem [\vel]
applies only to the gauge sector, whereas $\overline h_f(\mu)$ is of Higgs
origin.
For instance, $\Pi_{WW}(0)/M_W^2$ appears also in the expression for the
$\rho$ parameter, but there is also a similar contribution due to the $Z$
boson to cancel the term proportional to $M_H^2$.
At the other extreme, a Higgs boson with $M_H\ll2m_t$ does not produce
large corrections as may be seen from Eqs.~{\dbot}, {\dtau}, and {\heavyt}.

Equations~{\dtop}--{\dtau} have been obtained from the weak part of
Eq.~{\delf} by omitting subleading corrections.
Detailed inspection of the latter reveals that their $M_H$ dependence is
approximatively logarithmic, while they do not depend on $m_t$,
except for the subleading correction to $\delta_t^{\rm w}(\mu)$,
which, to good approximation, depends logarithmically on $m_t$.
We may exploit these observations by constructing very simple approximations
for the electroweak part of Eq.~{\delf}.
We find that $\delta_f^{\rm w}(m_f)+\delta_f^{\rm QED}(m_f)$ are very well
approximated by Eqs.~{\dtop}--{\dtau} if the terms
$$
d_f=a_f+b_f\ln{M_H\over300\gev}+c_f\ln{m_t\over175\gev}
\eqn\appr$$
are included on their right-hand sides.
The coefficients, $a_f$, $b_f$, and $c_f$, for the most important cases,
$f=t$, $b$, and $\tau$, are listed in the Table.
We shall illustrate the goodness of these approximations in the next section.

\medskip
\noindent
{\bf Table:}
Coefficients, $a_f$, $b_f$, and $c_f$, to be inserted in Eq.~{\appr} for
$f=t$, $b$, and $\tau$.
\medskip
\begintable
$f$ \| $a_f$ | $b_f$ | $c_f$ \crthick
$t$ \| $6.90\times10^{-3}$ | $1.73\times10^{-3}$ | $-5.82\times10^{-3}$ \cr
$b$ \| $1.52\times10^{-2}$ | $1.73\times10^{-3}$ | 0 \cr
$\tau$ \| $1.59\times10^{-2}$ | $1.73\times10^{-3}$ | 0
\endtable
\medskip

Since $\overline h_f(\mu)$ is a running parameter that satisfies the RG
equation,
$$
{\d\overline h_f(\mu)\over\d\ln\mu}=\beta_f\,,
\eqn\rge$$
we can, in general, expect to absorb all the LL corrections by a suitable
choice of $\mu$.
In particular, we expect the LL terms in $\delta_f(\mu)$ to vanish if we
choose $\mu=m_f$.
However, this is only true in a theory without any intermediate scale, \ie,
all other particle masses have to be small as compared to $\mu$.
Obviously, this is only true for $m_t$ when we assume that $M_H=\O(M_Z)$,
which we shall do henceforth.
In general, there are other particles, $X_i$ ($i=1,2,\ldots$), with masses
$M_i>m_f$ that are decoupled from the theory for $\mu<M_i$.
This induces in $\delta_f(\mu)$ LL corrections proportional to $\ln(M_i/\mu)$.
Clearly, this only affects $\delta_f^{\rm w}(\mu)$.

For a quantitative discussion of this issue, we rewrite Eq.~{\rge} as
$$
{\d\delta_f^{\rm w}(\mu)\over\d\ln\mu}={\beta_f^{\rm w}\over h_f}\,,
\eqn\delfw$$
where $\beta_f^{\rm w}=\beta_f+3h_f\left(Q_f^2\alpha+C_F\as\right)/(2\pi)$
is the weak part of Eq.~{\betaf}.
Next, we split $\beta_f^{\rm w}/h_f$ into a part proportional to $h_t$ and the
remainder, which, for simplicity, we associate with the scale $M_Z$, \ie,
$\beta_f^{\rm w}=\beta_f^t+\beta_f^Z$.
The effective running Yukawa coupling in the presence of decoupling is thus
determined by
$$
{\d\delta_f^{\rm eff}(\mu)\over\d\ln\mu}={1\over h_f}
\left[\beta_f^t\theta(\mu-m_t)+\beta_f^Z\theta(\mu-M_Z)\right]\,.
\eqn\delfeff$$
At sufficiently high scales, $M$, with $M\ge m_t$, this effective Yukawa
coupling coincides with the original one, so that
$\delta_f^{\rm w}(M)=\delta_f^{\rm eff}(M)$.
Solving Eqs.~{\delfw} and {\delfeff} and using this boundary condition, we
obtain, for arbitrary $\mu$,
$$
\delta_f^{\rm w}(\mu)=\delta_f^{\rm eff}(\mu)+\delta_f^{\rm LL}(\mu)\,,
\eqno\eq$$
where the LL terms are contained in
$$
\delta_f^{\rm LL}(\mu)=-{1\over h_f}\left[
\theta(m_t-\mu)\beta_f^t\ln{m_t\over\mu}
+\theta(M_Z-\mu)\beta_f^Z\ln{M_Z\over\mu}\right]\,.
\eqno\eq$$
Of course, $\delta_t^{\rm LL}(m_t)=0$.
For the reader's convenience, we list the corresponding threshold expressions
for $f=b$ and $\tau$:
$$\eqalign{
\delta_b^{\rm LL}(m_b)&=-{1\over16\pi^2}\left\{
h_t^2\left(N_c-{3\over2}\right)\ln{m_t\over m_b}
+\left[h_b^2\left(N_c+{3\over2}\right)+\sum_{i\ne t,b}N_ih_i^2-{9\over4}g^2
-{5\over12}g^{\prime2}\right.\right.\cr
&\qquad{}+\left.\left.\vphantom{\sum_{i\ne t,b}}
{2\over3}e^2\right]\ln{M_Z\over m_b}\right\}\,,\cr
\delta_\tau^{\rm LL}(m_\tau)&=-{1\over16\pi^2}\left[
N_ch_t^2\ln{m_t\over m_\tau}
+\left({5\over2}h_\tau^2+\sum_{i\ne\tau,t}N_ih_i^2-{9\over4}g^2
-{15\over4}g^{\prime2}+6e^2\right)\ln{M_Z\over m_\tau}\right]\,.\cr}
\eqn\llterms$$
We have seen above that a heavy Higgs boson does not decouple but
produces corrections proportional to $M_H^2$.
We have, therefore, refrained from separating out the logarithms involving
$M_H$.

In any RG analysis of the Yukawa couplings in GUT models,
one decouples the top quark  by setting $\overline h_t(\mu)=0$ for
$\mu<m_t$. The $W$ and $Z$ bosons can be decoupled
by replacing the $\beta$ functions of the unbroken
SU(2)$_L\otimes$U(1)$_Y$ theory with the ones of the
U(1)$_{\rm em}$ theory for $\mu<M_Z$.
This automatically takes care of the LL terms, which
are summed over by solving the RG equation numerically.
We have also separated out the QCD contribution, which is
universal, since the electroweak gauge and Higgs sectors do not interact
strongly.
We have proceeded similarly also for the purely electromagnetic contribution.

\chap{Numerical analysis}

We are now in a position to explore the phenomenological consequences
of our results.
To start with, we specify the input values for our numerical analysis.
We use $M_W=80.24$~GeV [\lep], $M_Z=91.1895$~GeV [\blo],
$m_\tau=1.777$~GeV [\wei], $m_b=4.72$~GeV [\dom], and
$m_t=(175\pm25)$~GeV [\cdf].
We adopt the other quark masses from Ref.~{\jeg}.
All other input parameters are taken from Ref.~{\pdg}.

In Fig.~{\figtwo}, we show the electroweak threshold corrections to
Eq.~{\defymsbar},
$\delta_f^{\rm w}(m_f)+\delta_f^{\rm QED}(m_f)$,
for (a) $f=t$, (b) $f=b$, and (c) $f=\tau$
at one loop in the SM as functions of $M_H$ assuming $m_t=(175\pm25)$~GeV.
Obviously, Eqs.~{\dtop}--{\dtau} plus Eq.~{\appr} lead to excellent
approximations in all cases throughout the full parameter space (dotted lines).
For $f\ne t$, decoupling effects are not corrected for, \ie, the results
still include the large, negative LL terms given in Eq.~{\llterms}.
Subtraction of these terms leads to Figs.~{\figthree}(a) and (b) for
$f=b$ and $\tau$, respectively.
Looking at Figs.~{\figtwo}(a), {\figthree}(a), and {\figthree}(b), we see
that, for $m_H = \O(M_Z)$, the threshold corrections are modest, well below
1\% in size.
Due to their quadratic dependence on $M_H$, they can become as large as 3\%
for $M_H=1$~TeV.
In addition, we have presented in Fig.~{\figthree}(c)
the difference of the results contained in Figs.~{\figthree}(a) and (b),
which is relevant in SU(5) GUTs for the $\tau$--$b$ Yukawa unification.
We find a negative effect, which is insensitive to $M_H$ and whose magnitude
increases with $m_t$.
However, even for $m_t=200$~GeV, it only amounts to approximately $-0.25\%$.

\chap{Conclusions}

We have calculated the full one-loop correction to the relation between the
running $\msbar$ Yukawa coupling, $\overline h_f(\mu)$,
and the pole mass, $m_f$, of a fermion in the SM
and shown that this correction is gauge independent.
As a by-product, we have introduced a gauge-independent definition of the
running $\msbar$ fermion mass, $\overline m_f(\mu)$.
In a way, our work complements the paper by Sirlin and Zucchini [\sirlin]
where the relation between the running $\msbar$ quartic coupling and the
pole mass of the Higgs boson has been evaluated at one loop in the SM.
Our result is particularly useful in the context of GUTs, where one wishes
to determine the implications of Yukawa-coupling unification at the GUT
scale for low-energy observables, which depend on the physical particle
masses.
To our knowledge, previous analyses have only included the familiar QCD
term in the quark case.
We found that the additional electroweak threshold corrections are relatively
modest, below 1\% in the low-$M_H$ region,
after the LL terms originating from the decoupling of weak physics at scales
$M_Z$ and $m_t$ have been subtracted.
At the other extreme, these threshold corrections have a quadratic high-$M_H$
behaviour and may reach values up to 3\% at $M_H=1$~TeV.

\vskip1.cm
\ACK
One of us (BAK) would like to thank William Marciano and Robin Stuart for
useful communications regarding the definition of the $\msbar$ fermion
mass, Elisabeth Kraus and Klaus Sibold for instructive discussions
concerning the formulation of the RG equations in the presence of masses,
and the KEK Theory Group for the warm hospitality extended to him during his
visit, when this work was finalized.
The work of BAK was supported by the Japan Society for the Promotion of
Science (JSPS) through Fellowship No.~S94159.
\endpage

\APPENDIX{A}{: One-loop amplitudes}

Here, we list the residual amplitudes that enter the evaluations of
$\overline m_f(\mu)$ and $\overline h_f(\mu)$ at one loop in the SM.
Having verified that these two quantities are gauge independent,
we may choose the 't~Hooft-Feynman gauge.
It is convenient to introduce the following short-hand notations:
$\Delta(m^2)=\Delta-\ln(m^2/\mu^2)$,
$w=M_W^2$, $z=M_Z^2$, $h=M_H^2$, $x=m_f^2$, $y=m_{f^\prime}^2$,
$v_f=\left(I_f-2\s^2Q_f\right)/(2\c\s)$, and $a_f=I_f/(2\c\s)$, where $I_f$
is the third component of weak isospin of $f_L$.

The sum of the vector and scalar components of the self-energy of
an on-mass-shell fermion, $f$, may be written as [\hff]
$$\eqalign{
\Sigma_V^f(x)+\Sigma_S^f(x)&={\alpha\over4\pi}\left\{
\left[{3\over8\s^2}\left(1+{x-y\over
w}\right)+{1\over8\c^2}-3\left(v_f^2-a_f^2+Q_f^2\right)\right]\Delta(x)
\right.\cr
&\qquad{}+{1\over4\s^2w}\left[{(x-y)^2+w(x+y)-2w^2\over2x}F(x,y,w)
+\left(2x-{h\over2}\right)F(x,x,h)\right.\cr
&\qquad{}+{(x-y)^3+4wx^2+3w^2(x+y)-2w^3\over4x^2}\ln{y\over w}
+\left({x-3y\over2}+w\right)\ln{x\over y}\cr
&\qquad{}+\left.{h\over2}\left(3-{h\over2x}\right)\ln{x\over h}
+4x-{5\over2}y+{w-h\over2}+{y^2+yw-2w^2\over2x}\right]\cr
&\qquad{}+\left[-v_f^2\left(2+{z\over x}\right)+a_f^2\left(4-{z\over x}\right)
\right]F(x,x,z)
+\left[-v_f^2{z^2\over2x^2}\right.\cr
&\qquad{}+\left.\left.a_f^2\left(2+3{z\over x}-{z^2\over2x^2}\right)\right]
\ln{x\over z}
-v_f^2\left(4+{z\over x}\right)+a_f^2\left(6-{z\over x}\right)-4Q_f^2\right\}
\,,\cr}
\eqno\eq$$
where
$$
F(x,y,z)=\cases{
\displaystyle
{\sqrt\lambda\over x}\arcosh{y+z-x\over2\sqrt{yz}}\,,
&if $x\le\left(\sqrt y-\sqrt z\right)^2$\cr
\displaystyle
-{\sqrt{-\lambda}\over x}\arccos{y+z-x\over2\sqrt{yz}}\,,
&if $\left(\sqrt y-\sqrt z\right)^2<x\le\left(\sqrt y+\sqrt z\right)^2$\cr
\displaystyle
-{\sqrt\lambda\over x}\left(\arcosh{x-y-z\over2\sqrt{yz}}-i\pi\right)\,,
&if $x>\left(\sqrt y+\sqrt z\right)^2$\cr}\,,
\eqn\fun$$
with $\lambda=x^2+y^2+z^2-2(xy+yz+zx)$.
For $x=y$, Eq.~{\fun} simplifies to
$$
F(x,x,z)=\cases{
\displaystyle
{z\over x}\sqrt{1-{4x\over z}}\,\arcosh\left({1\over2}\sqrt{z\over x}\,\right)
\,,
&if $\displaystyle x\le{z\over4}$\cr
\displaystyle
-{z\over x}\sqrt{{4x\over z}-1}\,\arccos\left({1\over2}\sqrt{z\over x}\,\right)
\,,
&if $\displaystyle x>{z\over4}$\cr}\,.
\eqno\eq$$
The bosonic and fermionic contributions to the self-energy of a $W$ boson
with zero invariant mass,
$\Pi_{WW}(0)=\Pi_{WW}^{\rm bos}(0)+\Pi_{WW}^{\rm fer}(0)$,
are given by [\hzz]
$$\eqalign{
\Pi_{WW}^{\rm bos}(0)&={\alpha w\over4\pi\s^2}\left[
\left(-2+{1\over\c^2}\right)\Delta(w)
+\left(2+{1\over\c^2}-{17\over4\s^2}\right)\ln\c^2
-{3\over4}\,{h\over w-h}\ln{w\over h}\right.\cr
&\qquad{}-\left.{17\over4}+{7\over8\c^2}-{h\over8w}\right]\,,\cr
\Pi_{WW}^{\rm fer}(0)&={\alpha\over8\pi\s^2}\sum_{(U,D)}N_{(U,D)}\left\{
-m_U^2\left[\Delta\left(m_U^2\right)+{1\over2}\right]
-m_D^2\left[\Delta\left(m_D^2\right)+{1\over2}\right]\right.\cr
&\qquad{}+\left.
{m_U^2m_D^2\over m_U^2-m_D^2}\ln{m_U^2\over m_D^2}\right\}\,,\cr}
\eqno\eq$$
where $(U,D)$ runs over all fermion doublets.
Finally, the tadpole contribution reads [\hzz]
$$\eqalign{
T&={e\over16\pi^2\s M_W}\left\{
w\left(3w+{h\over2}\right)\Delta(w)
+{z\over2}\left(3z+{h\over2}\right)\Delta(z)
+{3\over4}h^2\Delta(h)
+w^2+{z^2\over2}
\vphantom{\sum_i}\right.\cr
&\qquad{}
+\left.
{h\over2}\left(w+{z\over2}+{3\over2}h\right)
-2\sum_iN_im_i^4\left[\Delta\left(m_i^2\right)+1\right]\right\}\,,\cr}
\eqno\eq$$
where $i$ runs over all massive fermions.

\endpage
\refout
\bigskip
\figout
\end